\pdfoutput=1 % only if pdf/png/jpg images are used
\documentclass{JINST}

\title{The nucleation parameter for heavy-ion induced bubble nucleation in superheated emulsion detector.}

\author{Susnata Seth$^a$,\thanks{Corresponding author.}~
Mala Das$^a$, Sudeb Bhattacharya$^b$, Pijushpani Bhattacharjee$^{a,c}$ and  Satyajit Saha$^b$\\
\llap{$^a$}Astroparticle Physics and Cosmology Division, Saha Institute of Nuclear Physics,\\
  1/AF Bidhannagar, Kolkata 700064, India.\\
\llap{$^b$}Applied Nuclear Physics Division, Saha Institute of Nuclear Physics,\\
  1/AF Bidhannagar, Kolkata 700064, India.\\
\llap{$^c$}McDonnell Center for the Space Sciences \&
Department of Physics, Campus Box 1105,
Washington University in St. Louis,
One Brookings Drive,
St. Louis, MO 63130-4899. USA.
E-mail: \email{susnata.seth@saha.ac.in}}

\abstract{The values of the nucleation parameter,  \emph{k}, for bubble nucleation induced by high energy heavy ions 
$^{12}$C ($180$ MeV/u),  $^{20}$Ne ($400 $ MeV/u) and $^{28}$Si ($350 $ MeV/u) in superheated emulsion detector are determined by 
comparing the experimentally obtained normalized count rates with those obtained from simulations done using the GEANT3.21 simulation code.
The results show that the nucleation parameter depends on the mass number of the incident heavy ions, and decreases with increasing 
mass number.}


\begin{document}
\section{Introduction}
A superheated emulsion detector consists of a large number of drops of   superheated 
liquid suspended in another immiscible liquid-like, soft gel medium or a firm polymer matrix. One form of this detector is known as 
``superheated drop detector (SDD)'' \cite{bib1}
and another form is ``bubble detector (BD)''\cite{bib2}.  The basic principle of operation of these 
 detectors is same as that of a bubble chamber \cite{bib3}.
During the passage of an energetic particle through a drop, if the energy deposited within a certain critical 
length is larger than a certain critical energy,
the superheated liquid undergoes a phase transition to vapour phase, \it i.e.\rm, a  nucleation event occurs.
Acoustic pulse generated in this process constitutes the signal, which is recorded by acoustic sensors. Since 1979 \cite{bib1}, the 
superheated emulsion is being used in various areas such as in neutron dosimetry 
\cite{bib4,bib5,bib6,bib7}, gamma ray detection \cite{bib7,bib8,bib9}, proton detection \cite{bib10}, neutron spectrometry 
\cite{bib11,bib12,nsp5,nsp7} and also in cold dark matter 
 search experiments \cite{w4,w5,w6}. 

An important parameter that characterizes the bubble nucleation process in superheated emulsion detectors is the nucleation parameter. 
There are various phenomenological  models to describe the nucleation process in superheated emulsion \cite{r5,npa2}.
In the present paper, we use the nucleation parameter \emph {k} as defined in ref. \cite{r5}, discussed  in  
more details in  Section \ref{sec:cobf}. There have been a large number of studies which attempt to estimate the value of nucleation 
parameter for neutrons \cite {bib7,r5,npa2,npa3,npb1,npb2}.
But not much work has been done on estimation of nucleation parameter for heavy ions. In an earlier work \cite{h4},
the response of SDD to various high energy heavy ions has been 
studied and the nucleation parameter has been estimated from the experimental results using estimation of dE/dx
from SRIM2008 code \cite{b3}. In the present work, we consider the detailed geometry
 of the detector, allowing us to make a more accurate estimate of dE/dx, resulting in a more reliable estimate of the 
nucleation parameter for heavy ions.
 
Experimental studies on detection of heavy ions by superheated emulsion detectors have been done in refs.
\cite{h4,h1,h2,h3}. From  measurement of the maximum track length for various ions in a  superheated emulsion,
 the relationship between maximum track
length and the atomic number of the ions can be established, allowing identification of heavy ions \cite{h1,h2,h3}. 
The high energy heavy ions used in these studies were  $^{12}$C (290 MeV/u), $^{28}$Si (600 MeV/u),
  $^{40}$Ar (650 MeV/u), $^{56}$Fe (500 MeV/u), $^{84}$K (400 MeV/u) and   $^{132}$Xe (290 MeV/u). The response, threshold 
temperatures and the threshold degree of metastability of nucleation in superheated emulsion made of active liquid  R-114 were 
investigated for different 
heavy ions with different energies, namely, $^{12}$C (180 MeV/u, 230 MeV/u), $^{20}$Ne (230 MeV/u, 400 MeV/u),
 $^{28}$Si (180 MeV/u, 490 MeV/u), $^{40}$Ar (500 MeV/u), and
$^{56}$Fe (500 MeV/u) \cite{h4,h5}. The track length for protons has also been studied and in this case
 bubble tracks are formed by protons in 
the region corresponding to the Bragg peak of protons in the detector \cite{bib10}. 

To understand the response of superheated emulsion detectors to various particles, one has to take recourse 
 to  simulation. Computational studies of the response of superheated emulsion to neutrons  
have been done using the frame work of  Seitz's radiation-induced nucleation theory \cite{sim1,b1}. 
Monte Carlo studies of the response of   superheated drops of liquid C$_4$F$_{10}$ (b.p.: $ -1.7 ^{\rm o} $C) for alpha particles,
neutrons, gamma rays 
and $\delta$-rays have been done within the context of 
dark matter search experiments \cite{w4,sim3}. These studies have allowed determination of various parameters like 
 the alpha detection  efficiency  as a function of temperature and the loading factor of the detector.
It was also established from simulation studies that the maximal alpha detection 
efficiency is  inversely proportional to the drop radius.

In the present work, we present our results of  the simulation of 
the response of superheated emulsion with 
active liquid R-114 (C$_2$Cl$_2$F$_ 4 $;  b.p. $3.77^\circ$ C)  to heavy ions 
 $^{12}$C ($180 $ MeV/u), $^{20}$Ne ($400 $ MeV/u) and $^{28}$Si ($350 $ MeV/u)
 using GEANT3.21 simulation toolkit \cite{b2}. 
We also study the possible dependence of the nucleation parameter  \emph{k} \cite{r5} on the mass
of the heavy ions.
We have performed two separate sets of  simulations, referred to as  Simulation I and Simulation II.
 In Simulation I, we  simulated the  geometry of  the experimental set up of the actual experiment
to calculate the expected nucleation rate.
 In Simulation II, instead of simulating the whole experimental set up, we independently determined 
the bubble nucleation probability of a single drop as a function of the nucleation 
parameter, \emph{k}, using GEANT3.21. This nucleation probability is used
to calculate the expected number of nucleation events for the
same experiment. The normalized count rates at 
the threshold temperature of bubble nucleation, for the high energy heavy ions $^{12}$C ($180 $ MeV/u), 
$^{20}$Ne ($400 $ MeV/u) and $^{28}$Si ($350 $ MeV/u) from the above two simulations are compared
 with the experimental data \cite{h4} to estimate the nucleation parameter \emph{k}, for the heavy ions.
\footnote{Lest there be confusion, we mention here that the results of the present paper are not directly
 applicable to the case of WIMP (cold dark matter candidate) search using SDDs. This is because of the huge
 difference in the energy scales of the ions causing the nucleation in the two cases. In the WIMP case,
 the WIMP-induced recoil nuclei causing the bubble nucleation events are expected to have kinetic energies of
 order tens to hundreds of keV, whereas here we are concerned with nucleations caused by heavy ions with energies
 of order hundreds of MeV per nucleon.}

\section{Condition of bubble formation}\label{sec:cobf}

The critical energy, W, for bubble nucleation at a particular temperature and pressure as given by 
Gibb's reversible thermodynamics \cite{gibbs}, can be expressed as 
\begin{equation}
 \rm W=\frac{ \rm 16 \pi \sigma ^3(T)}{\rm 3(P_v-P_0)^2} \; ,
\end{equation}

where $\sigma(\rm T)$ is the surface tension at temperature T, P$\rm _v$ is the vapour pressure of the liquid at the temperature T 
and P$\rm _0$
is the ambient pressure. The term $\rm (P_v-P_0)$ is the degree of superheat of the liquid at the temperature T.
 If the radius of the embryo is greater than a certain 
critical radius ($ \rm r_{c}$), it grows till the whole liquid drop is vaporized. 
If on the other hand, the radius of the vapour embryo is smaller than the critical radius, it shrinks back to the liquid.
The critical radius is given by , 
\begin{equation}
 \rm r_{c}=\frac{\rm 2\sigma (T)}{\rm P_v-P_0} \; .
\end{equation}

Usually, a very small fraction of the total deposited energy ($\rm E_{dep}$) is utilized in the nucleation process.
The quantity $\frac{\rm W}{\rm E_{dep}}$ is often called 
as  the thermodynamic efficiency ($\eta_{\rm T}$) for bubble nucleation process \cite{b4}.
 The value is typically in the range of 3\% to 5\% for the neutron 
induced nucleation \cite{b4}. 
The condition for  bubble formation can be written as,
\begin{equation}
\label{eq.1} 
\frac{\rm W}{\emph{k}\, \rm r_{c}}(\rm T,P)=\frac{\rm dE}{\rm dx}\; ,
\end{equation}
 where, \emph{k} is the nucleation parameter \cite{r5} and  is 
equal to twice the thermodynamic efficiency ($\eta_{\rm T}$) \cite{bib12}. The study of the nucleation parameter (\emph{k}) 
has been described in details in the literature \cite{bib12,r5,h4,b6}. From the experiment, the estimated value of \emph{k}
with neutrons from Am-Be source was found to be $0.1158$ for R-114 \cite{b6}. For the heavy ions the estimated
value of  \emph{k} for the same liquid is $0.11$ \cite{h4}.

%%%%%%%%%%%%%%%%%%%%%%%%%%%%%%%%%%%%%%%%%%%%%%%%%%%%%%%%%%%%%%%%%%%%%%%%%%%%
\begin{figure}
\centering
\includegraphics[width=.65\textwidth]{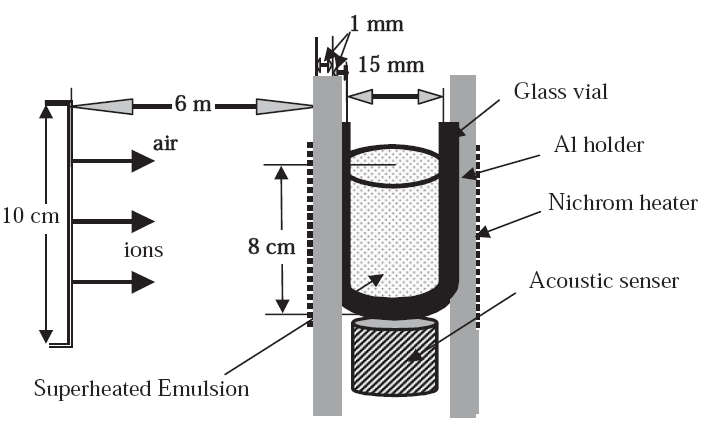}
\caption{Experimental set up.}
\label{setup} 
\end{figure}

%%%%%%%%%%%%%%%%%%%%%%%%%%%%%%%%%%%%%%%%%%%%%%%%%%%%%%%%%%%%%%%%%%%%%%%%%%%%%%%%%%

\section{Present work}

The details of the experiment  for which the simulation is carried out is described in 
ref. \cite{h4}. A brief description of the experiment 
is given below for the sake of completeness. The experimental set up is shown in figure \ref{setup}.
 The superheated emulsion was inside a glass vial of 1 mm thickness. 
The vial was placed inside an aluminium holder of thickness 1 mm. The beam entry port was 6 m away from
 the detector system.
The dimensions of the detector used in the experiment were about $80$ mm in length and $15$ mm in diameter,
 containing $5000$ drops of the liquid  R-$114$ 
suspended in a firm polymer matrix. The drops had a distribution in diameter  with a sharp peak at about 21 $\mu$m.
 The main component of the polymer was glycerine. 
 Before  entering into the detector, the ions passed through $6$ m of air, $1$ mm of Al holder and $1$ mm of glass vial. 
In the experiment, the particle  flux ($\rm I_B$) was $1000$ particles/sec/cm$^{2}$. The simulation is carried out for 
$^{12}$C ($180 $ MeV/u), $^{20}$Ne ($400 $ MeV/u) and $^{28}$Si ($350 $ MeV/u).

%%%%%%%%%%%%%%%%%%%%%FIGURE
\begin{figure}
\centering
\includegraphics[width=.65\textwidth]{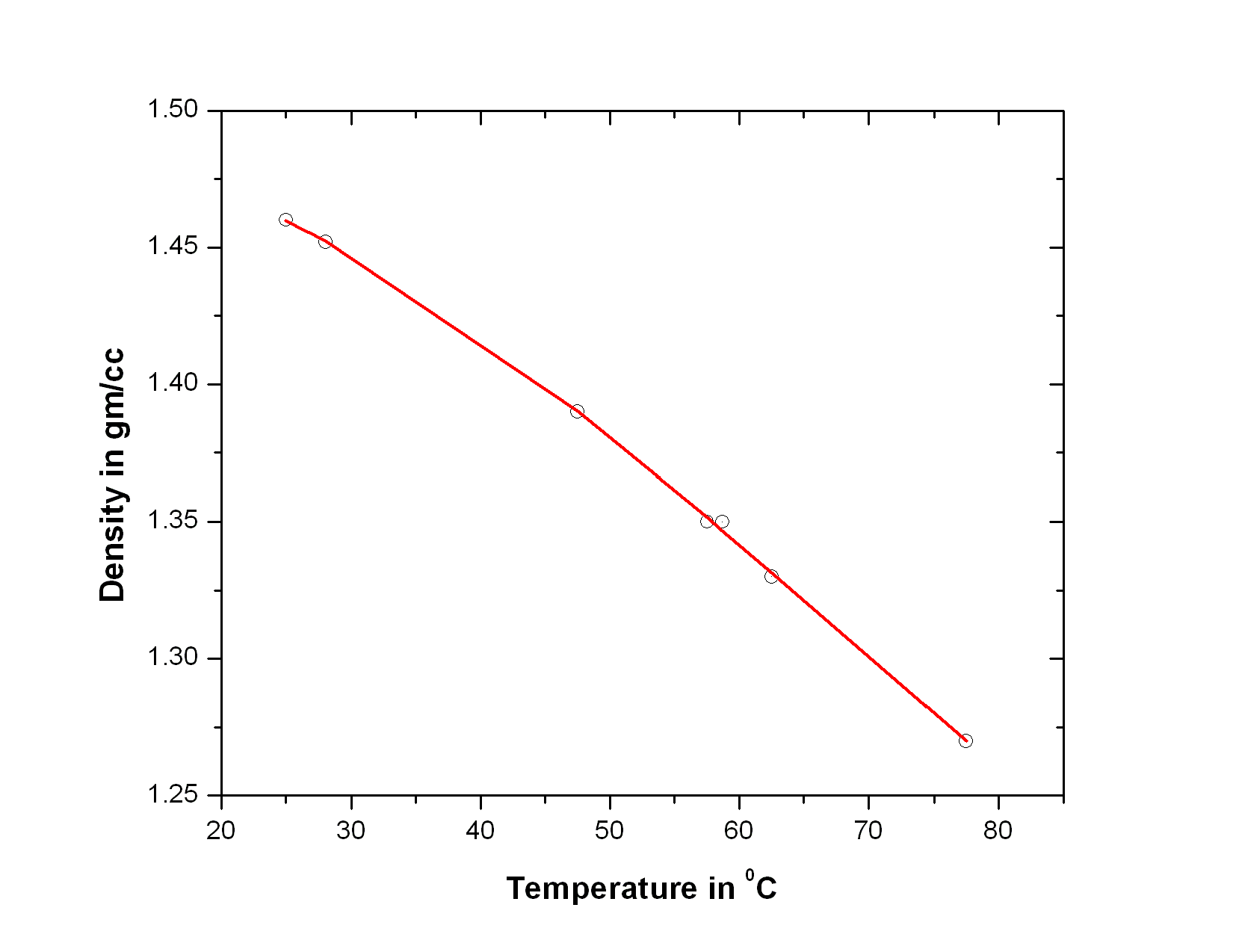}
\caption{ Variation of density vs temperature for R-114 liquid. }
\label{density}
\end{figure}

%%%%%%%%%%%%%%%%%%%%%%%%%%%%%%%%%%%%%%%%%%%%%%%%%%%%%%%%%%%%%%%%%%%%%%%%%%%%%%%%%%
\subsection{Simulation I }
\begin{figure}
\centering
\includegraphics[width=.45\textwidth]{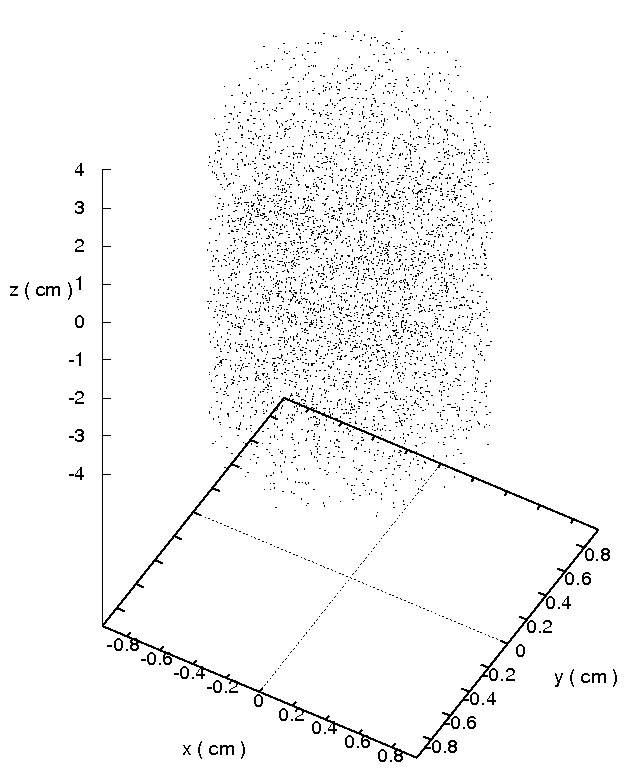}
\caption{Distribution of center of drops of R-114 in Simulation I.}
\label{drop} 
\end{figure}

In this simulation, we simulate the experimental set up of the actual experiment.
 The positions of the drops and incident points of the ions are chosen randomly.
There are two parts in the simulation:  the geometry and the tracking of particles. In the geometry part,
 first the glycerine cylinder of density $1.24$ gm/cc, glass cylinder of density $2.58$ gm/cc and the liquid drops of 
R-114 are considered. 
The densities of  R-114 liquid at different temperatures, are obtained from the following fitted functional form, 
shown in red line in figure \ref{density}, of the density as a function of temperature:
\begin{equation}
\label{dens_fit}
\rm y = 1.49059 + 8\times 10^{-5}\rm x - 6 \times 10^{-5}\rm x^{2} + 2.8291\times10^{-7}\rm x^{3} \;.
\end{equation}
Here y is density in gm/cc and x is temperature in $^{\circ}$C. 
The densities at some specific temperatures as shown in black circles in figure \ref{density} are obtained from ref. \cite{h4}.
A cylinder of Al with height 80 mm and diameter 19 mm is created. Within this Al cylinder,
 a 1 mm thick hollow glass cylinder of same height and outer diameter of 17 mm is placed. Then the cylinder of glycerine
with height 80 mm and diameter 15 mm is created and placed inside the hollow glass cylinder. A total
 of 5000 ($\rm N_{0}$)  uniformly and randomly distributed non-overlapping drops of R-114 are then placed in the cylinder containing
 the glycerine (shown in figure \ref{drop}).
The diameter of each drop is taken as 20 $\mu$m.

\begin{figure}
\centering
\includegraphics[width=.45\textwidth]{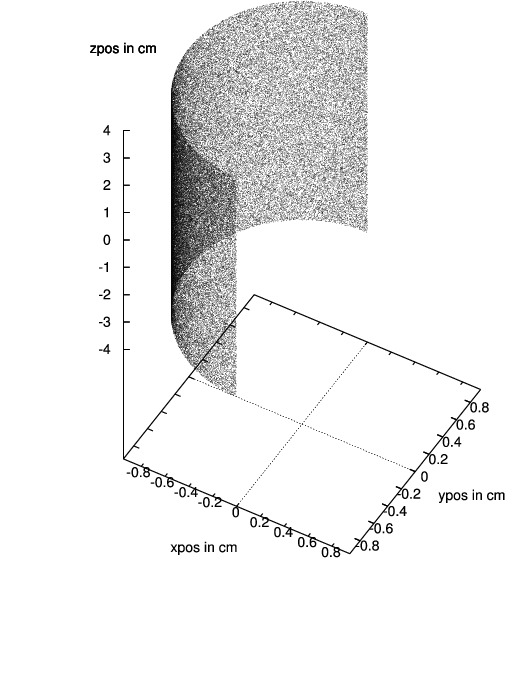}
\caption{Distribution of 2.6 million of incident positions of ions on the outer surface of the Al cylinder 
in Simulation I (For clarity, only 0.1 million incident positions are shown).
Note that the random distribution of the incident points spans almost a  continuous surface.}
\label{part} 
\end{figure}

The second part of the simulation is tracking of particles through the geometry described above.
The energy deposition by the ions during the passage through 6 m of air is obtained using SRIM 2008 code \cite{b3}.
 As the simulation is done for an experiment of 100 sec time duration, a total of  $\sim$ $2.6$ million particles 
coming from negative x-direction is allowed to fall randomly on the outer  surface of the Al cylinder perpendicular to the long axis of 
the detector. All particles have momenta only in the positive x-direction and the values of momenta
at the entry point of the Al cylinder for 
$^{12}$C ($180 $ MeV/u), $^{20}$Ne ($400 $ MeV/u) and $^{28}$Si ($350 $ MeV/u)
are found to be 7.071 GeV/c, 18.759 GeV/c and 24.022 GeV/c respectively.
The distribution of the incident positions of the ions is shown in figure \ref{part}.
The distribution of the drop centers and incident positions are generated outside the GEANT code,
using the SRAND random number generator \cite{r8}.

 The Linear Energy Transfer, dE/dx, within a drop is obtained from the GEANT code.
 A true bubble nucleation event is assumed to occur if dE/dx is greater than the critical energy
deposition ($\frac{\rm W}{k\rm r_{c}}$) required for nucleation.
Extra events due to more than one nucleation for same drop are rejected. The normalized count,
 $\rm \frac{\rm 1}{\rm N_{0}.I_{B}}(\frac{\rm dN}{\rm dt})$ in unit of $\rm cm^2$,  is calculated and plotted as a
function of operating temperature. 
%%%%%%%%%%%%%%%%%%%%%FIGURE

%%%%%%%%%%%%%%%%%%%%%%%%%%%%%%%%%%%%%%%%%%%%%%%%%%%%%%%%%%%%%%%%%%%%%%%%%%%%%%%%%%

%%%%%%%%%%%%%%%%%%%%%%%%%%%%%%%%%%%%%%%%%%%%%%%%%%%%

%%%%%%%%%%%%%%%%%%%%%%%%%%%%%%%%%%%%%%%%%%%%%%%%%%%%%%%%%%%%%%%%%%%%%%%%%%%%%%%%%%%%

\subsection{Simulation II }

This simulation is carried out  to first find the probability of nucleation of a drop for the three ions, 
 $^{12}$C ($180 $ MeV/u), $^{20}$Ne ($400 $ MeV/u) and $^{28}$Si ($350 $ MeV/u) and then, using this probability, 
the normalized count rate and the nucleation parameter for the above mentioned experiment  are obtained.
In this simulation only one single drop of liquid R-114 of diameter 20 $\mu$m is created. 
To include the actual  experimental situation of the randomness in the incident positions of the ions  within a drop, 
a total of 
$100$ incident positions within the drop are generated using the SRAND random number generator \cite{r8}.  
 On each incident position an ion with same energy is made to fall $3000$ times.
The energy losses of the ions during their
passage through air are taken into account and calculated using SRIM 2008. As a result, the momenta of  
$^{12}$C ($180 $ MeV/u), $^{20}$Ne ($400 $ MeV/u) and $^{28}$Si ($350 $ MeV/u) at the incident points within the drop 
are 7.071 GeV/c, 18.759 GeV/c and 24.022 GeV/c respectively. The momentum is only in  the positive x-direction. 
%%%%%%%%%%%%%%%%%%%%%%%%%%%%%%%%%%%%%%%%%%%%%%%%%%%%%%%%%%%%%%%%%%%%%%%%%%%%%%%%%%
\begin{figure}
\centering
\includegraphics[width=.65\textwidth]{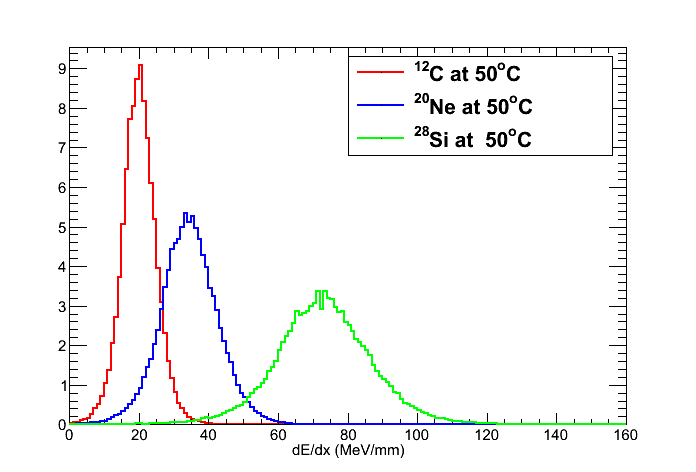}
\caption{ Distribution of LET within a 20 $\mu$m diameter R-114 drop obtained at 
50$^{\rm o}$C from GEANT3.21 for the simulation of $^{12}$C ($180 $ MeV/u), 
$^{20}$Ne ($400 $ MeV/u) and  $^{28}$Si ($350 $ MeV/u).
The red, blue and green curves are for $^{12}$C (180 MeV/u), $^{20}$Ne ($400 $ MeV/u)
 and  $^{28}$Si ($350 $ MeV/u) with momenta 7.071 GeV/c, 18.759 GeV/c and 24.022 GeV/c, respectively. }
\label{LETdist}
\end{figure}
%%%%%%%%%%%%%%%%%%%%%%%%%%%%%%%%%%%%%%%%%%%%%%%%

\begin{figure}
\centering
\includegraphics[width=.75\textwidth]{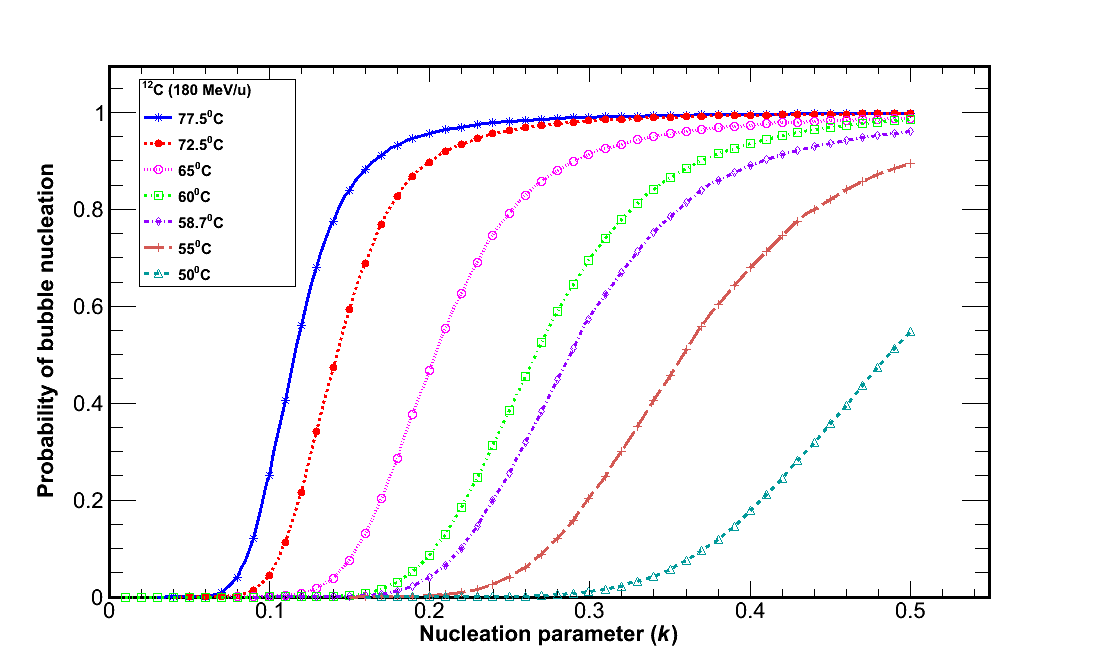}
\caption{Probability of bubble nucleation as a function of nucleation parameter {\emph k} 
for $^{12}$C (180 MeV/u) with incident momentum 7.071 GeV/c.}
\label{cmulti} 
\end{figure}

 The energy deposition as well as LET 
of the ion along its track through the drop are obtained from the GEANT code. The distribution of LET for the ion
has been obtained  for each temperature.
 In figure \ref{LETdist} the distribution of LET within a 20 $\mu$m diameter drop for  $^{12}$C,  $^{20}$Ne and $^{28}$Si 
with above mentioned momenta are shown for a temperature of  $50^{\rm o}$C. The
density of R-114 liquid  at different temperatures are obtained from the fitted functional form using eq. (\ref{dens_fit}).
Using the distribution of LET,
 the probability of bubble nucleation of a drop has been calculated for each temperature
for different values of  nucleation parameter (\emph{k}).
We defined the probability of bubble nucleation of a drop (P$_{B}$) for a particular nucleation parameter (\emph{k}) 
as,
\begin{equation}
 \rm {P_{\rm B}} =\frac{\rm No.\: of \: hits \;with  \; LET \;greater\; than\; \frac{W}{kr_c}}{Total\; number \;of \; hit}\; .
\end{equation}
It is seen that, for any given temperature, the probability reaches a saturation value 
as the value of  \emph{k}  is increased. For illustration, this probability P$_{\rm B}$ as a function of the nucleation parameter 
\emph{k}, for various temperatures, is shown in figure \ref{cmulti} for the heavy ion
 $^{12}$C (180 MeV/u) with incident momentum 7.071 GeV/c.
To obtain the number of bubble nucleation events due to a particular ion  with a particular 
nucleation parameter, the probability of bubble nucleation at that value of the nucleation parameter for that ion is multiplied by 
the total hit points  obtained from Simulation I for the same ion.  Finally the normalized count,
 $\rm \frac{1}{N_{0}.I_{B}}(\frac{dN}{dt})$ in unit of $\rm cm^2$,  is calculated and plotted as a
function of temperature.

%%%%%%%%%%%%%%%%%%%%%%%%%%%%%%%%%%%%%%%%%%%%%%%%%%%%%%%%%%%%%%%%%%%%%%%%%%%%%%%%%%%%
%%%%%%%%%%%%%%%%%%%%%%%%%%%%%%%%%%%%%%%%%%%%%%%%%%%%%%%%%%%%%%%%%%%%%%%%%%%%%%%%%%%%%%
 
\section{Results and Discussions}

\subsection{Simulation I }
From Simulation I,  we have obtained the normalized count rates for different temperatures. Varying the value of \emph{k}, 
the response has been fitted with the experimental result. 
To find the  value of \emph{k}, the normalized count rate ($\rm N_{\rm expt}$) at the experimentally
estimated threshold temperature and that obtained from the Simulation I (N$\rm_{1}$)  are compared.
Experimentally at low temperatures the response was almost flat at a value of $5 \times 10^{-7}$ cm$^{2}$ corresponding
to background. Above a certain threshold temperature, the number of counts rises sharply \cite{h4}. 
The experimental threshold temperature (T$_{\rm th})$ is considered at
the midpoint between the temperature at the end of the above mentioned flat region of the curve  
and the temperature corresponding to the next  higher count above the background count of  $5 \times 10^{-7}$ cm$^{2}$.
 The normalized count rate at the threshold temperature so defined is taken to be the mean of 
the two counts at the two corresponding temperatures.
 The same procedure is adopted here to find the normalized count rate (N$\rm_{1}$)
at the threshold temperature.

%%%%%%%%%%%%%%%%%%%%%%%%%%%%%%%%%%%%%%%%%%%%%%%%%%%%%%%%%%%%%%%%%%%%%%%%%%%%%%%%%%%
\begin{table}
 \caption { Normalized count rate obtained from simulations by varying the nucleation parameter (\emph{k}) and
compared with experimentally obtained values.}
\label{table1}
\smallskip
\centering
\begin{tabular}{|llllllc|}
\hline
&\multicolumn{1}{c}{Experiment}&&\multicolumn{2}{c}{Simulation I} &\multicolumn{2}{c}{Simulation II}\vline\\
\cline{2-2} \cline{4-7}
 Ion (energy     &Normalized   &  Value of  & Normalized  & $\Delta_{1}^2$ &   Normalized & $\Delta_{2}^2$\\
 in Mev/u)&count at T$_{\rm th}$  &      \emph{k}              & count  at T$_{\rm th}$    &    &                         count at T$_{\rm th}$  & \\
and  &  ($10^{-7}$  cm$^{2}$)&             &     cm$^{2}$          &      &                          cm$^{2}$          &                \\
 T$_{\rm th}$($^{\circ}$C)&  ( N$_{\rm expt}$ )&   &  ( N$_{1}$ ) &       &                           ( N$_{2}$ )         &                    \\

\hline

$^{12}$C (180)&$5.45$                  & 0.24   & $6.8\times 10^{-7}$& $0.061$  &  0  $7.68\times 10^{-7}$ &0.167     \\
58.7$\pm$1.2&      & 0.23   &$3.93\times 10^{-7}$& $0.078$      &   $5.61\times 10^{-7}$ &0.001    \\
               &      & 0.22   & $3.41\times 10^{-7}$& $0.140$&  $3.9\times 10^{-7}$ &0.081\\ 
               &     & 0.21    & $6.3\times 10^{-8}$& $0.782$&  $2.54\times 10^{-7}$ &0.285\\  
               & & 0.19& $7.0\times 10^{-9}$& $0.974$&  $8.72\times 10^{-8}$  &0.706\\
               &   & 0.11 &0                & $1.000$& $7.64\times 10^{-11}$ &1.000\\
\hline
$^{20}$Ne (400)&$8.1$& 0.23 & $3.126\times 10^{-6}$&$8.175$  &$3.60\times 10^{-6}$ &11.864\\
62.5$\pm$2.5& &0.12 & $1.551\times 10^{-6}$& $0.837$& $1.09\times 10^{-6}$ &0.119\\
&& 0.11 & $2.8\times 10^{-7}$&$0.428$   &$6.29\times 10^{-7}$ &0.050\\
&& 0.10 & $1.2\times 10^{-7}$&$0.726$ &$2.83\times 10^{-7}$& 0.423\\
\hline
$^{28}$Si (350 )&$5.6$& 0.23&$3.192\times 10^{-6}$& $22.090$&$3.81\times 10^{-6}$ &33.681\\
57.5$\pm$2.5&& 0.08&$1.598\times 10^{-6}$&  $3.436$&$1.39\times 10^{-6}$ &2.197\\
&& 0.07&$8.32\times 10^{-7}$& $0.236$&$4.90\times 10^{-7}$ &0.016\\
&& 0.06&$1.2\times 10^{-8}$& $0.958$& $8.67\times 10^{-8}$ &0.714\\
&& 0.05&$1.00\times 10^{-9}$& $0.996$&$7.91\times 10^{-9}$ &0.972\\
\hline
\end{tabular}

\end{table}

%%%%%%%%%%%%%%%%%%%%%%%%%%%%%%%%%%%%%%%%%%%%%%%%%%%%%%%%%%%%%%%%%%%%%%%%%%%%%%%%%%%%
\begin{figure}[tbp]
\centering
\includegraphics[width=.75\textwidth]{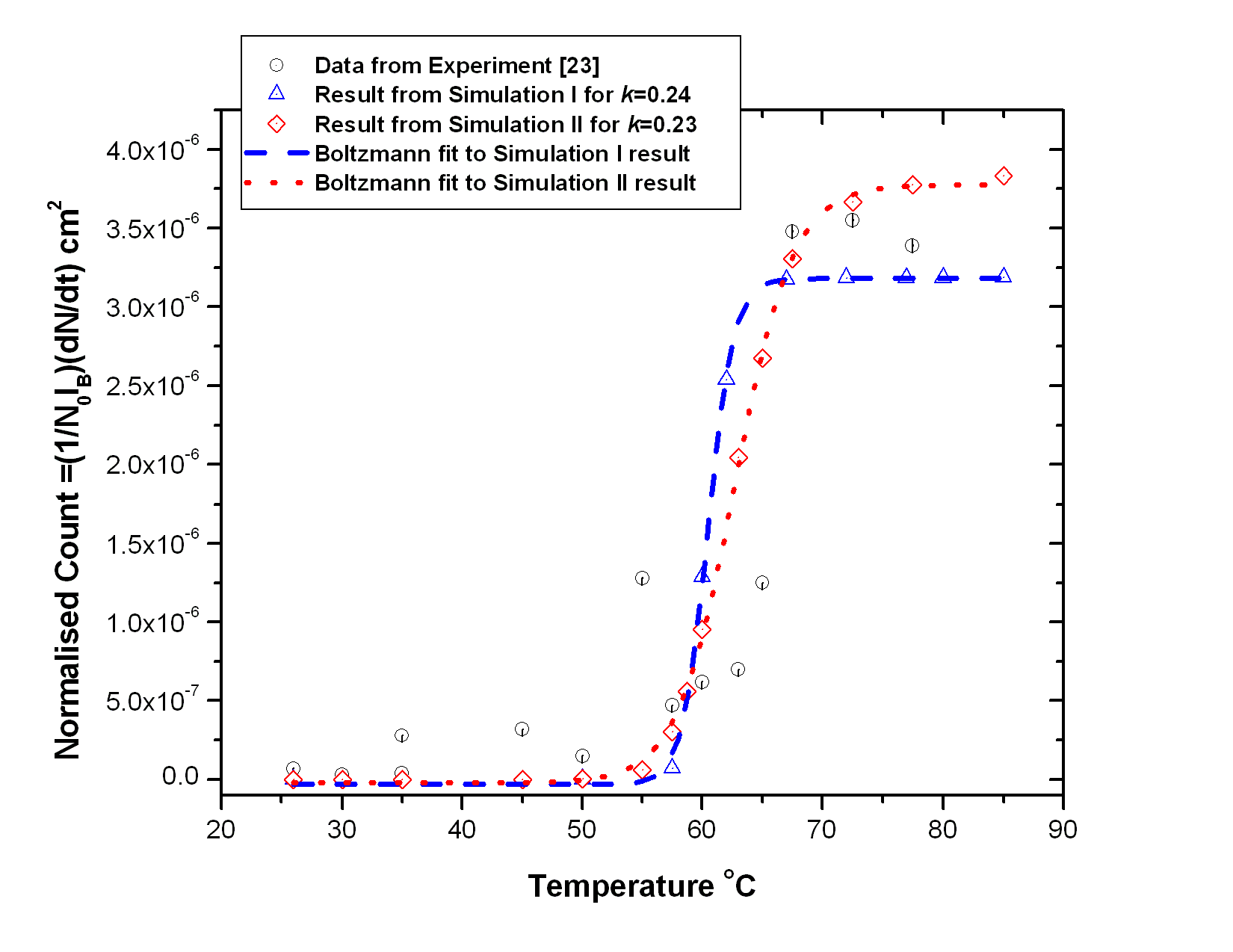}
% Here is how to import jpg/ png/ pdf figure 
%if you are using pdflatex command to compile
\caption{Simulation results for  $^{12}$C ($180 $ MeV/u) ion.}
\label{fig1}
\end{figure}
%%%%%%%%%%%%%%%%%%%%%%%%%%%%%%%%%%%%%%%%%%%%%%%%%%%%%%%%%%%%%%%%%%%%%%%%%%%%%%%%%%%%

\begin{figure}
\centering
% Here is how to import EPS figure if you are using latex command to compile
\includegraphics[width=.75\textwidth]{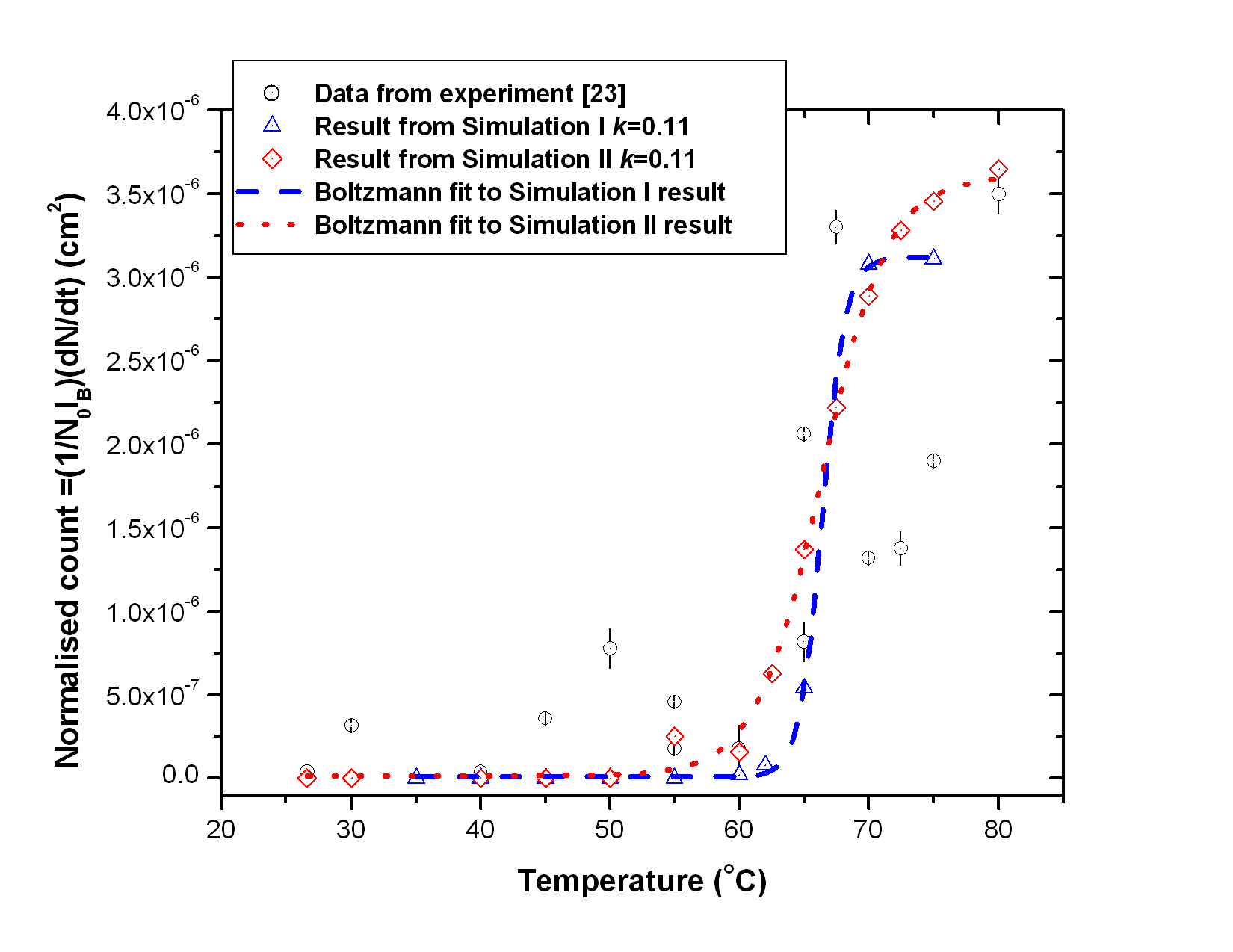}
% Here is how to import jpg/ png/ pdf figure 
%if you are using pdflatex command to compile
\caption{Simulation result for  $^{20}$Ne ($400 $ MeV/u) ion.}
\label{figNe}
\end{figure}

In table {\ref{table1}}, the normalized count rates at the threshold temperature from  
the experiment and Simulation I with different values of \emph{k} are tabulated. The value of \emph{k} 
for which the deviation 
between experiment and Simulation I is least is taken as best values of  \emph{k} for a particular ion.
The deviation for Simulation I is defined as $\Delta_{1}^2=( 1- \frac{\rm N_{1}}{\rm N_{\rm expt}})^2$.
In case of $^{12}$C ($180 $ MeV/u), the deviation ($\Delta_{1}^2$)  is smallest for nucleation parameter \emph{k} = 0.24. 
For  the other two heavier ions, $^{20}$Ne ($400 $ MeV/u) and $^{28}$Si ($350 $ MeV/u),
the deviation is smallest for the values of nucleation parameter 0.11 and 0.07 respectively.
The normalized count rate as a function of temperatures for the best value of \emph{k} obtained above for Simulation  I 
(blue triangle  in figures \ref{fig1}, \ref{figNe}, \ref{figSi})
are fitted to the Boltzmann function (blue dashed line in figures \ref{fig1}, \ref{figNe}, \ref{figSi} ) given by
 \[ \rm y(T)=\frac{(A_1-A_2)}{(1+e^{\frac{(T-T_o)}{ dT}})}+A_2 \; ,\] where y(T) is the
normalized count at temperature (T$^{\rm o}\rm C$) and $\rm A_1$ is the base line value and 
$\rm A_2$ is the plateau value of the count rate. 
 The values of the thermodynamic efficiency, $\eta_{T}$ for  $^{12}$C ($180 $ MeV/u), $^{20}$Ne ($400 $ MeV/u) 
and $^{28}$Si ($350 $ MeV/u)
from Simulation I  is 0.12, 0.055 and 0.035, respectively (table \ref {tab:table2}).

%%%%%%%%%%%%%%%%%%%%%%%%%%%%%%%%%%%%%%%%%%%%%%%%%%%%%%%
\begin{table}[tbp]

\caption{ Nucleation parameter (\emph{k}) and thermodynamic efficiency ($\eta_{\rm T}$) 
for different heavy ions in R-114 liquid obtained from Simulation I and Simulation II.}
\label{tab:table2}
\smallskip
\centering
\begin{tabular}{|llllllc|}
\hline
%&&& \center{2}{\emph{k} obtained from}  &\center{2}{ $\eta_{\rm T}$ obtained from \emph{k} }\\
Ion & Incident & Mass no. &\multicolumn{2}{c}{\emph{k} obtained from}& \multicolumn{2}{c}{$\eta_{\rm T}$ obtained from \emph{k}}\vline\\
 \cline {4-5}\cline{6-7}
& energy & & &&&\\
& MeV/u& &Simulation I & Simulation II &Simulation I& Simulation II\\
\hline
$^{12}$C& $180$    & 12 & 0.24&0.23 & 0.12&0.115\\

$^{20}$Ne& $400$  & 20 & 0.11&0.11& 0.055&0.055\\

$^{28}$Si& $350$  & 28 & 0.07&0.07& 0.035&0.035\\
\hline
\end{tabular}
\end{table}
%%%%%%%%%%%%%%%%%%%%%%%%%%%%%%%%%%%%%%%%%%%%%%%%%%%%%%%

%%%%%%%%%%%%%%%%%%%%%%%%%%%%%%%%%%%%%%%%%%%%%%%%%%%%%%%%%%%%%%%%%%%%%%%%%%%
%%%%%%%%%%%%%%%%%%%%%%%%%%%%%%%%%%%%%%%%%%%%%%%%%%%%%%%%%%%%%%%%%%%%%%%%%%%

\subsection{Simulation II }
\begin{figure}
\centering
% Here is how to import EPS figure if you are using latex command to compile
\includegraphics[width=.75\textwidth]{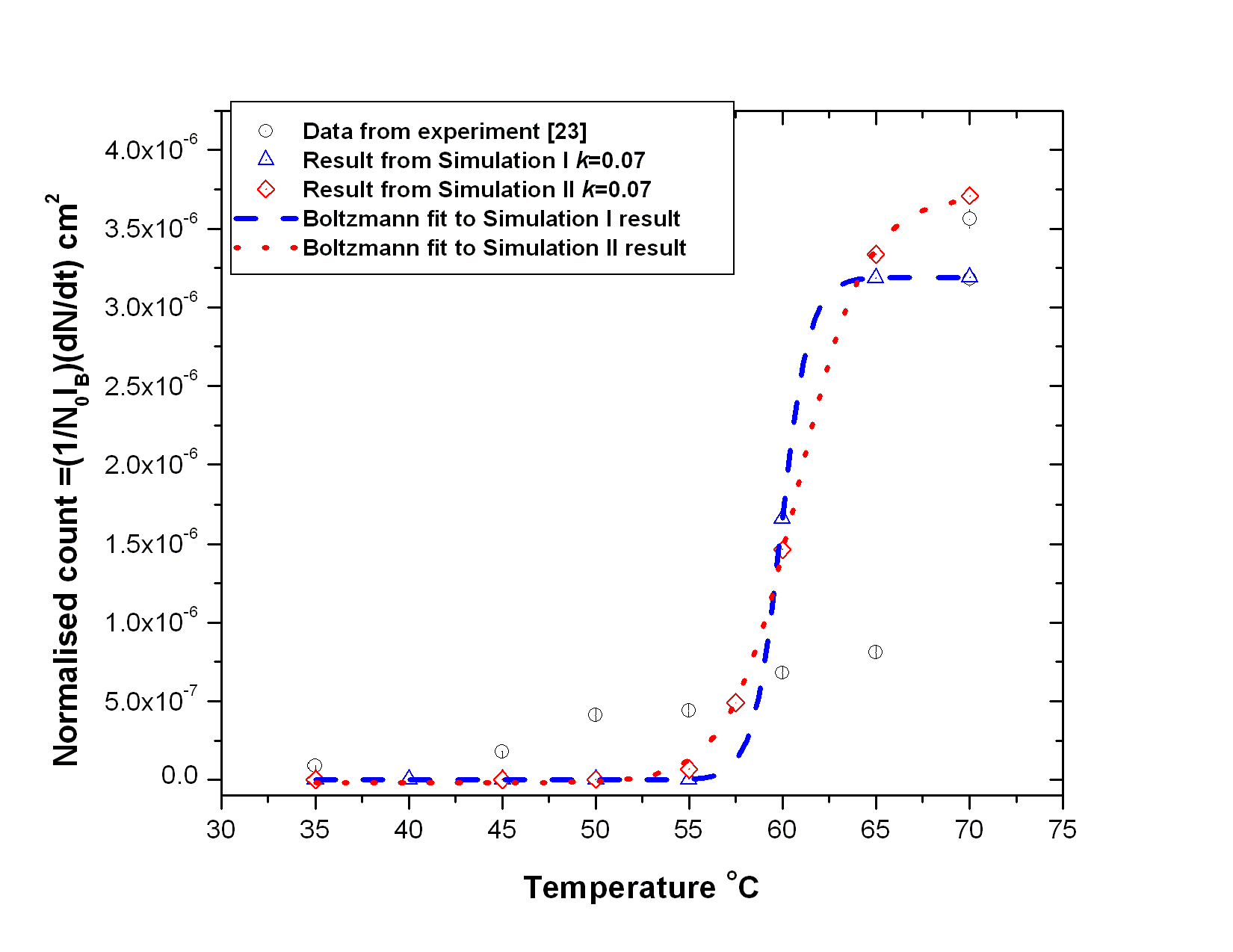}
% Here is how to import jpg/ png/ pdf figure 
%if you are using pdflatex command to compile
\caption{Simulation result for  $^{28}$Si ($350 $ MeV/u) ion.}
\label{figSi}
\end{figure}

A similar procedure  as described above for Simulation I is followed to obtain the nucleation parameter (\emph{k}) and
 thermodynamic efficiency ($\eta_{\rm T}$) from Simulation II.
The normalized count rates ($\rm N_2$) for different values of nucleation  parameter (\emph{k}) at the threshold temperature obtained 
from the simulation are shown in table {\ref{table1}}. The deviation is calculated using the eq. $\Delta_{2}^2=( 1- \frac{\rm N_{2}}{\rm N_{\rm expt}})^2$.
It is seen that the smallest deviation $\Delta_{2}^2$ is obtained for \emph{k} values of 
0.23, 0.11, 0.07 for  $^{12}$C ($180$ MeV/u), $^{20}$Ne ($400 $ MeV/u) and $^{28}$Si ($350$ MeV/u) respectively 
(table {\ref {table1}}).
The response obtained from Simulation II  
for $^{12}$C ($180$ MeV/u), $^{20}$Ne ($400 $ MeV/u) 
and $^{28}$Si ($350$ MeV/u)  ions are shown as red dotted line in figures \ref{fig1}, \ref{figNe}, \ref{figSi}, respectively.
 The thermodynamic efficiency ($\eta_{T}$) for the same set of ions are 0.115, 0.055 and 0.035, respectively (table \ref {tab:table2}).

%%%%%%%%%%%%%%%%%%%%%%%%%%%%%%%%%%%%%%%%%%%%%%%%%%%%%%%%%%%%%%%%%%%%%%%%%
%%%%%%%%%%%%%%%%%%%%%%%%%%%%%%%%%%%%%%%%%%%%%%%%%%%%%%%%%%%%%%%%%%%%%%

\subsection{General observations }
From Simulation I and II, $\eta_{\rm T}$ is found to be in the range of 3.5\% to 12\%. 
The values of \emph{k} and $\eta_{T}$ in R-114 liquid for different heavy ions as obtained from Simulation I and II 
are shown in table {\ref {tab:table2}}.
% For the present simulation $dE\over dx$ from ion-electronic interactions are more relevant than ion-nuclear interactions.
From table {\ref {tab:table2}}, comparing the value of  \emph{k} from the two simulations, it is seen that for  
 $^{28}$Si ($350 $ MeV/u) and  $^{20}$Ne ($400 $ MeV/u), the
values of  \emph{k} are same and these are 0.07 and 0.11, respectively, while for $^{12}$C ($180 $ MeV/u)  values of 
 \emph{k}
are 0.24 and 0.23 from Simulation I and Simulation II, respectively.
Our results show that the value of \emph{k} 
varies with the mass number of the ions.
It can be explained by comparing the  mean value of LET  distribution within a 20 $\mu$m drop obtained from the GEANT code.
The mean value of LET   for $^{28}$Si ($350 $ MeV/u) and  $^{20}$Ne ($400 $ MeV/u) are  
73.14 MeV/mm and 34.50 MeV/mm respectively. These values  are greater than that for  $^{12}$C ($180 $ MeV/u) (20.11 MeV/mm); 
(see figure \ref{LETdist}). 
By definition $\eta_{T}$ is the fraction of energy deposition required for 
bubble nucleation and as  $\eta_{T}$ decreases with increase of LET, this implies that  thermodynamic efficiency ($\eta_{T}$) is 
lowest for $^{28}$Si and highest for  $^{12}$C.

%%%%%%%%%%%%%%%%%%%%%%%%%%%%%%%%%%%%%%%%%%%%%%%%%
\begin{table}[tbp] % figures (and tables) should go top or bottom of
                    % the page where they are first cited or in
                    % subsequent pages
\caption{ Nucleation parameter (\emph{k}) 
for different heavy ions in R-114 liquid obtained using simulation I and all experimental data for C, Ne, Si.}
\label{tab:table3}
\smallskip
\centering
\begin{tabular}{|llllc|}

\hline
Ion & Incident energy & Mass no. & \emph{k}  & Deviation obtained \\
& MeV/u & &&using eq. (\ref{dev}) \\
\hline
$^{12}$C& $180$    & 12 & 0.19 &$0.725$ \\

$^{20}$Ne& $400$  & 20 & 0.11& $0.784$ \\

$^{28}$Si& $350$  & 28 & 0.05&$0.783$ \\
\hline
\end{tabular}
\end{table}
%%%%%%%%%%%%%%%%%%%%%%%%%%%%%%%%%%%%%%%%%%%%%%%%%%%%%%%%

We have also done a separate calculation to obtain the value of nucleation parameter 
using the result from Simulation I. In this calculation  
all the experimental results for the whole temperature range used in the experiment and for all
the three ions C, Ne and Si are considered. Varying the value of \emph{k} and
 using the experimental data
for all the three ions, a  best fitted  single simulation curve has been independently  found for each of the three ions,  
$^{12}$C ($180 $ MeV/u), $^{20}$Ne ($400 $ MeV/u) and $^{28}$Si ($350$ MeV/u).
The  deviation between the experimental and the
simulated  normalized count rate  has been calculated for all experimental data using the following formula,
\begin{equation}
\label{dev}
 \rm {deviation} =\frac{1}{N}\left [ \sum_{i=1}^{N}\left (1- \frac{N_{i}^{sim}}{N_{i}^{expt}}\right )^2 \right ],
\end{equation}
where, i represents the number of experimental data, N is the total number of experimental 
data available for  the three ions, obtained from \cite{h4}, N$_{\rm i}^{\rm expt \rm}$ and N$_{\rm i}^{\rm sim \rm}$
are the normalized count rate corresponds to a particular temperature from experiment and Simulation I respectively.
The best fitted single simulation curve for each ion corresponds to a nucleation parameter {\emph k} for which the 
deviation is smallest.
 For $^{12}$C ($180 $ MeV/u), 
$^{20}$Ne ($400 $ MeV/u) and $^{28}$Si ($350$ MeV/u)  
the nucleation parameter ( \emph{k}) is found to be equal to 0.19, 0.11 and 0.05, respectively (table \ref{tab:table3}), 
which also shows the dependence of 
\emph{k} on the mass number of the ions.

We should mention that, as evident from the figures above, there are some deviations of the simulation results from the 
actual experimental data points. This is only to be expected. There can be several reasons for this. For example, in the 
simulations described above, we have not  included the effects of possible fluctuation of the flux of the incident particles 
during the experiment. Also, we have considered all the drops to be of a fixed size, rather than a possible distribution of the 
radii of the drops in the actual experimental situation. Furthermore,  possible spontaneous nucleations due to impurities in the 
real detector are not considered.  All these effects may, to varying degrees, contribute to the deviations between the simulated 
results and experimental data points seen in the figures above. However, the important point to note is that the simulations do 
reproduce the broad nature of the experimental response curve. This, we believe, indicates that our simulations do indeed capture 
the essential physics of the nucleation process under consideration.
\section{Conclusions}
 We have studied the response of superheated emulsion detector to incident heavy ions,
  $^{12}$C ($180 $ MeV/u), $^{20}$Ne ($400 $ MeV/u) and  $^{28}$Si ($350 $ MeV/u) to understand the dependence of the
bubble nucleation process on the masses of the incident nuclei. We have done this by simulating the energy losses (LET)
by the nuclei during their passage through the active liquid drops that constitutes the detector.
Our simulation also includes the geometry of the detector used in the actual experiment. We have determined the nucleation
parameter, \emph{k}, by comparing the simulation results with those from experiment.
 It is observed from the result that the value of the nucleation parameter, \emph{k} for $^{12}$C ($180 $ MeV/u) is higher 
than that of the higher mass ions like $^{20}$Ne ($400 $ MeV/u) and  $^{28}$Si ($350 $ MeV/u). For the three ions investigated here, 
\emph{k} is observed to decrease with the increase in mass of the heavy ions. The present simulation provides an 
important observation that the nucleation parameter, (\emph{k}) depends on the mass number of the ions.

\acknowledgments
The authors would like to thank Sunanda Banerjee for valuable discussions
and Lab Saha for his help in software installation. One of us (PB) wishes
to thank Ramanath Cowsik for support under the Clark Way Harrison Visiting
Professorship program at the McDonnell Center for the Space Sciences at
Washington University in St. Louis.

\end{document}